\documentclass[a4paper,11pt]{article}
\pdfoutput=1
\usepackage{jcappub}
\usepackage{journal_abbrev}
\usepackage{xparse,verbatim}
\usepackage{xspace,ulem}
\xspaceaddexceptions{]\}}
  \usepackage{bm}
  \usepackage{hyperref}

\newcommand{\onefigs}[3]{
  \begin{figure}%
    \centerline{\resizebox{0.5\textwidth}{!}{\includegraphics*{#1}}}%
    \caption{#3}\label{#2}%
  \end{figure}%
}
\newcommand{\twofig}[4]{%
  \begin{figure*}%
    \centerline{\resizebox{\hsize}{!}{\includegraphics*{#1} \,%
        \includegraphics*{#2}}}%
    \caption{#4}\label{#3}%
  \end{figure*}%
}

\newcommand{\sect}[1]{Sec.~\ref{#1}\xspace}

\newcommand{\app}[1]{Appendix~\ref{#1}\xspace}

\newcommand{\eq}[1]{Eq.~(\ref{#1})\xspace}
\DeclareDocumentCommand{\eqs}{m m m o o}{%
  \IfNoValueTF {#4} {%
    Eqs.~(\ref{#1}){\xspace #2} (\ref{#3})%
  }{%
    Eqs.~(\ref{#1}){\xspace #2} (\ref{#3}){\xspace #4} (\ref{#5})%
  }%
}

\newcommand{\equ}[2][]{\begin{equation}\label{#1}#2\end{equation}}
\newcommand{\eqa}[2][]{\begin{align}\label{#1}#2\end{align}}
\newcommand{\eqm}[2][]{\begin{multline}\label{#1}#2\end{multline}}
\newcommand{\se}{\ensuremath{\sigma_\mathrm{8}}\xspace}
\newcommand{\hse}{\ensuremath{\hat{\sigma}_\mathrm{8}}\xspace}

\newcommand{\nn}{\nonumber\\}
\newcommand{\specialcell}[2][c]{%
  \begin{tabular}[#1]{@{}c@{}}#2\end{tabular}}

\newcommand{\pprof}{\ensuremath{P^\mathrm{prof}}\xspace}

\newenvironment{referee}{\bf}{}
\newcommand{\bref}{\begin{referee}}
\newcommand{\eref}{\end{referee}}

\def\ba#1\ea{\begin{align}#1\end{align}}
\def\bea{\begin{eqnarray}}
\def\eea{\end{eqnarray}}
\def\be{\begin{equation}}
\def\ee{\end{equation}}
\def\d{\delta}
\def\s{\sigma}
\def\({\left(}
\def\){\right)}
\def\[{\left[}
\def\]{\right]}

\def\<{\left\langle}
\def\>{\right\rangle}
\def\lapl{\nabla^2}

\def\ohat#1{#1}

\newcommand{\vs}{\nonumber\\}
\def\d{{\delta}}
\def\eps{{\varepsilon}}

\renewcommand{\v}[1]{\bm{#1}}

\def\vx{\v{x}}

\def\vk{\v{k}}

\def\vp{\v{p}}

\def\O{\mathcal{O}}

\def\clapl{c_{\lapl\d}}

\def\fsum#1{\sum_{#1\neq 0}^{k_{\rm max}}}

\def\Mpch{\,h^{-1}\text{Mpc}}
\def\iMpch{\,h\,\text{Mpc}^{-1}}
\def\Msunh{\,h^{-1} M_\odot}
\def\Plin{P_\text{L}}
\def\Pol{P_\text{1-loop}}
\def\Ptl{P_\text{2-loop}}

\def\Lbox{L_\text{box}}
\def\L{\Lambda} 

\newcommand{\refeq}[1]{Eq.~(\ref{eq:#1})}

\newcommand{\reffig}[1]{Fig.~\ref{fig:#1}}

\newcommand{\reftab}[1]{Tab.~\ref{tab:#1}}
\newcommand{\refsec}[1]{Sec.~\ref{sec:#1}}

\newcommand{\refapp}[1]{Appendix~\ref{app:#1}}

\def\emph#1{\textit{#1}}

\title{Cosmology Inference from a Biased Density Field using the EFT-based Likelihood}

\author[a]{Franz Elsner,}
\author[a]{Fabian Schmidt,}
\author[b]{Jens Jasche,}
\author[c]{Guilhem Lavaux,}
\author[a]{and Nhat-Minh Nguyen}

\emailAdd{felsner@mpa-garching.mpg.de}
\emailAdd{fabians@mpa-garching.mpg.de}
\emailAdd{jens.jasche@fysik.su.se}
\emailAdd{guilhem.lavaux@iap.fr}
\emailAdd{minh@mpa-garching.mpg.de}

\affiliation[a]{Max--Planck--Institut f\"ur Astrophysik,
  Karl--Schwarzschild--Stra\ss e 1, 85748 Garching, Germany}
\affiliation[b]{The Oskar Klein Centre, Department of Physics, Stockholm University, Albanova University Center, SE 106 91 Stockholm, Sweden}
\affiliation[c]{Sorbonne Universit\'e, CNRS, UMR 7095, Institut d'Astrophysique de Paris, 98 bis bd Arago, 75014 Paris, France}

\keywords{cosmic web, cosmological parameters from LSS, redshift surveys}

\arxivnumber{1234.56789}

\abstract{
  The effective-field-theory (EFT) approach to the clustering of galaxies and other biased tracers allows for an isolation of the cosmological information that is protected by symmetries, in particular the equivalence principle, and thus is robust to the complicated dynamics of dark matter, gas, and stars on small scales. All existing implementations proceed by making predictions for the lowest-order $n$-point functions of biased tracers, as well as their covariance, and comparing with measurements.
  Recently, we presented an EFT-based expression for the conditional probability of the density \emph{field} of a biased tracer given the matter density field, which in principle uses information from arbitrarily high order $n$-point functions. Here, we report results based on this likelihood by applying it to halo catalogs in real space, specifically an inference of the power spectrum normalization \se. We include bias terms up to second order as well as the leading higher-derivative term. For a cutoff value of $\L = 0.1\iMpch$, we recover the ground-truth value of \se to within 95\% CL for different halo samples and redshifts.
  We discuss possible sources for the remaining systematic bias in \se as well as future developments.
}

\begin{document}

\maketitle

\flushbottom

\section{Introduction}
\label{sec:intro}

State-of-the-art approaches for the analysis of large-scale structure
(LSS) data typically make use of summary statistics like the two-point
correlation function to compare theoretical models to observational
data \cite[e.g.,][]{1969PASJ...21..221T, 1974ApJS...28...19P,
  1977ApJ...216..682F, 1978MNRAS.182..673P, 1986ApJS...62..301S,
  1996MNRAS.283..709H, 2001ApJ...554..857G, 2002ApJ...579...42C,
  2003ApJ...595...59B, 2005ApJ...619..697A, 2005ApJ...633..560E,
  2007ApJ...658...85M, 2011ApJ...736...59Z,
  2017arXiv170801530D}. However, as has been
realized early on \cite{1934ApJ....79....8H},
the galaxy distribution is non-Gaussian, and even an
infinite hierarchy of higher order correlation functions may be
insufficient to fully capture all cosmological information encoded in
LSS data sets \cite{1991MNRAS.248....1C, 2011ApJ...738...86C}.

Alternative approaches have since been developed that take a
different, more ambitious avenue to cosmological signal
inference. Instead of focusing on summary statistics, they aim
directly at reconstructing the three-dimensional underlying matter
density field from observations of astrophysical tracers like galaxies
\cite{1989ApJ...336L...5B, 1994ApJ...423L..93L, 1995MNRAS.272..885F,
  1999AJ....118.1146S, 2004MNRAS.352..939E, 2010MNRAS.406...60J,
  2010MNRAS.407...29J, 2010MNRAS.409..355J, 2010MNRAS.403..589K,
  2012MNRAS.420...61K, 2013MNRAS.432..894J, 2015MNRAS.446.4250A,
  2017MNRAS.467.3993A}. In an important step,
Ref.~\cite{2013MNRAS.432..894J} improved on previous implementations by
introducing a physical forward model for the matter field as a means to more precisely
predict the intricate statistical properties of the evolved density
field. Starting from a set of simple Gaussian initial conditions at
high redshift as probed by cosmic microwave background radiation
experiments \cite[e.g.,][]{2003ApJS..148..175S, 2007ApJS..170..377S,
  2009ApJS..180..330K, 2011ApJS..192...18K, 2014A&A...571A..16P,
  2016A&A...594A..13P, 2016A&A...596A.107P}, nonlinear effects of
gravitational collapse are taken into account via approximate
analytical or purely numerical methods to compute the corresponding
evolved density field at low redshift that can then be compared to
observations. Owing to the Gaussian nature of the initial density
field, the two-point correlation function becomes a lossless form of
data compression in this regime and fully captures all information
relevant to cosmology. As a by-product of extending the reconstruction
to the initial density field, a fully probabilistic description of all
possible histories of the structure formation process compatible with
the analyzed data set becomes available, offering interesting
opportunities to implement other cosmological tests
\cite[e.g.,][]{2018arXiv180807496K}.

Unsurprisingly, these sophisticated approaches are faced with numerous
challenges that are both theoretical and numerical in nature. For
example, exploring an extremely high-dimensional parameter space
typical of the reconstruction of a three-dimensional field
necessitates the use of sophisticated sampling algorithms
\cite{Duane1987}, \cite[e.g.,][]{2010ApJ...724.1262E,
  2010MNRAS.407...29J}. Further, since a comprehensive theoretical
understanding of the formation of actual observed tracers such as
galaxies is still lacking, an effective
model for galaxy biasing is necessary to connect any underlying dark
matter density field to actual observables like galaxy positions and
redshifts \cite{1984ApJ...284L...9K, 1986ApJ...304...15B,
  1989MNRAS.237.1127C, 1993ApJ...413..447F, 1996MNRAS.282..347M,
  1999MNRAS.308..119S, 2018PhR...733....1D}. Finally, a statistical
comparison of observational data to theoretical predictions requires
the use of a likelihood function that quantifies the probability to
observe the given data realization under the model
hypothesis.

Numerous functional forms for this likelihood have been proposed to
compare model predictions to observations
\cite[e.g.,][]{1956AJ.....61..383L, 1984ApJ...276...13S,
  1994ApJ...423L..93L, senatore:2015, MSZ}.
  Based on the \emph{effective field theory
  (EFT)} approach \cite{baumann/etal:2012,carrasco/etal:2012},
however, only recently a rigorous
description for the LSS likelihood has been developed \cite{paperI,cabass/schmidt:2019}. Most notably, and in contrast to previous
attempts at modeling the statistics of LSS data, this likelihood is
formulated in Fourier space. In this paper, we use numerical
simulations to provide an in-depth analysis of the EFT LSS
likelihood, using dark matter halos in different mass bins as tracers.
In particular, we test whether it allows for unbiased cosmological parameter inference
from reconstructed matter density fields.

Since they are based on simulations, our tests use the correct
phases of the initial density field. Hence, we eliminate an important
source of uncertainty that would normally result in noticeably larger error
bars on cosmological parameters in real surveys. Our tests are thus more
stringent to pass compared to a realistic application, where the initial density field is unknown and has to be simultaneously inferred from the data.

The paper is organized as follows. In \sect{sec:like} we briefly
review the statistical framework to analyze LSS data developed in
\cite{paperI}, which forms the basis of our analysis. We also present
an analytical marginalization scheme which greatly reduces the number
of free parameters that need to be varied.
We then introduce a set of simulations used to assess the performance of
the likelihood in \sect{sec:sim}. After presenting our implementation
in \sect{sec:impl}, we turn to the results in \sect{sec:results}.
We summarize our findings in \sect{sec:concl}.
The appendices present further details and a derivation of the Poisson
expectation for stochasticity.

\section{Method}
\label{sec:like}

We start by giving an overview of the main aspects of the methodology
used in our analysis, and then introduce a useful extension, analytical marginalization, that aims at obtaining numerical results more efficiently.

\subsection{Recap: Fourier-space likelihood}
\label{sec:likeF}

In this section, we provide a short summary of the results presented
in \cite{paperI}. Our main objective is to obtain a
probabilistic reconstruction of the initial density field
$\vec{\d}_\text{in}$, cosmological parameters, and additional nuisance
parameters necessary to capture the uncertainties of the process of
structure formation. To do so, we need to model the effects of various
aspects of the underlying physical processes. More precisely, we have
to specify a prior characterizing the statistical properties of the
initial density perturbations, and provide a forward model, describing
the gravitational collapse of matter overdensities over
time. Additionally, we need to include a bias model, connecting the
observed tracers to the matter density field, and, finally, a
likelihood, describing the stochastic aspects involved in this process
and therefore allowing for a quantitative comparison of model predictions
to observations.

We begin by reviewing the basic notation introduced in
\cite{paperI} that we will make use of in the remainder
of this paper. For simplicity and in view of the tests presented in
\sect{sec:s8profile}, we will refer to the tracers of the matter
density field as halos. Then, given the evolved matter density field
$\vec{\d}$, at any order in perturbation theory, the deterministic halo
density, $\vec{\d}_{h, \mathrm{det}}$, can be expressed as linear sum
over a finite set of operators, $\vec{O}$,
\equ[eq:dhdet]{
  \vec{\d}_{h, \mathrm{det}}[\vec{\d}, \{ b_O \}] = \sum_O b_O
  \vec{O}[\vec{\d}] \, ,
}
where the $b_O$ are the associated bias parameters (see
\cite{2018PhR...733....1D} for a review). Here and throughout, the
vector notation denotes fields, in our case discretized on a uniform cubic grid.
In the following, we will use the discrete Fourier transform, so that fields in Fourier space are dimensionless as well. 
Following Ref.~\cite{paperI}, we use the bias expansion up to second
order in perturbations, i.e., we restrict ourselves to the following set of operators:
\equ[eq:operators]{
  O \in \{ \d, \ \lapl\d, \ \d^2, \ K^2 \} \, ,
}
  where $\d$ is the fractional matter density perturbation, and
  \be
K^2 \equiv (K_{ij})^2 = \left( \left[\frac{\partial_i\partial_j}{\lapl} - \frac13 \d_{ij} \right] \d \right)^2
  \ee
  is the tidal field squared. The corresponding bias parameters are denoted as
\equ{
  \mathcal{B}_O
    = \{ b_1, \ \clapl, \ b_2/2, \ b_{K^2} \} \, .
}
The coefficient of $\lapl\delta$ is denoted as $\clapl$ to emphasize
that it also absorbs other higher-order contributions, as discussed in \cite{paperI}. 
All operators in \refeq{operators} are constructed from the density
field filtered with a sharp low-pass filter in Fourier space which only
retains Fourier modes at $|\vk| < \Lambda$, where $\Lambda$ functions as a
cutoff for the theoretical description.
  The sharp cut in Fourier space naturally arises in effective field theory
  approaches which integrate out initial density perturbations above a cutoff $\Lambda$ \cite{cabass/schmidt:2019}, although this does not mean that other filter shapes are excluded in principle. However, we derived in \cite{paperI} that a sharp-$k$ filter is a condition for obtaining an unbiased maximum-likelihood point of the Fourier-space likelihood in the form used here.

Moreover, the operators in
\refeq{dhdet} are renormalized, which in this case essentially means
that the quadratic operators $\d^2$ and $K^2$ are made to be orthogonal
to $\d$. We refer to
\app{app:operators} for details on the exact definition and
renormalization procedure. It is worth noting that the renormalization
is not essential to obtain an unbiased \se estimate, but allows for
a comparison of the resulting bias parameters with measurements from
$n$-point functions.

Notice that \refeq{dhdet} involves an expansion in two
small parameters, essentially orders of perturbations and spatial derivatives
(see Sec.~4.1 of \cite{2018PhR...733....1D} for a detailed discussion).
Here, we assume that both small parameters are comparable, which leads us
to include terms up to second order in perturbations as well as the
leading higher-derivative operator $\lapl\d$ in \refeq{operators}. The
reasoning behind this is discussed in greater detail in \cite{paperI}. 

As derived in \cite{paperI}, for an observed halo field,
$\vec{\d}_{h}$, we can then compute the conditional probability for the halo density field given the matter density in Fourier space,
\equ[eq:PcondGFT]{
  \ln P\left(\vec{\d}_h\Big|\vec{\d}, \{b_O\}\right) =
  -\fsum{\vk} \left[ \frac{1}{2} \ln \s^2(k) + \frac{1}{2 \s^2(k)}
   \left| \d_h(\vk) - \d_{h,\rm det} [\vec{\d}, \{b_O\}](\vk) \right|^2 \right] \, .
}
In the following, we will refer to this conditional probability simply as ``likelihood,'' since it is the part of the overall likelihood of biased tracers that is relevant for the study presented in this paper. 
Here, $\s^2(k)$ is a dimensionless variance that is analytic in $k$, specifically a power series in $k^2$. For our numerical implementation, we include terms up to order $k^4$, and parametrize $\s^2(k)$ as follows:
\equ[eq:noise_var]{
  \s^2(k)
  = \left( \s_{\eps} + k^2 [ \s_{\eps,2} + b_1 \s_{\eps\eps_m,2} ] \right)^2.
}
The parametrization is chosen so that $\s^2(k)$ is positive definite.
$\s_\eps^2$ can be interpreted as the amplitude of halo stochasticity in the large-scale limit ($k \to 0$). In \app{app:variance} we derive the expectation for $\s_\eps^2$ for a Poisson process, which will be useful for the interpretation of numerical results on this parameter. $\s_{\eps,2}^2$ is the leading scale-dependent correction to the halo stochasticity (see Sec.~2.7 of \cite{2018PhR...733....1D} for a discussion). 
Finally, $\s_{\eps\eps_m,2}$ captures the cross-correlation of stochasticity
in the halo and matter fields; the stochasticity in the matter field $\eps_{m}(\vk)$ is constrained by mass- and momentum conservation to be of order $k^2$ on large scales.

The closed-form expression for the likelihood in \eq{eq:PcondGFT}
affords a straightforward way to derive maximum likelihood estimates
for bias parameters, and, as shown in \cite{paperI}, for
the cosmological parameter $\s_8$. 
This framework was used in \cite{paperI} to demonstrate
unbiased parameter estimation from LSS data without the explicit use
of conventional summary statistics. Here, we will obtain results using the
field-level likelihood, rather than the analytical maximum-likelihood point discussed in \cite{paperI}. 

The degeneracy between $b_1$ and \se, which is perfect in linear theory,
is broken when including nonlinear information. In particular, the fact that the displacement term contained in the second-order
matter density is also multiplied by $b_1$, coupled with the fact that the second-order
matter density scales differently with \se than the linear-order one.
Thus, fundamentally, the
possibility of estimating \se in this way is due to the equivalence principle,
which ensures that galaxies move on the same trajectories as matter on
large scales, and thus requires that the second-order displacement term is multiplied
by the same bias coefficient as the linear-order density field (see also
Sec.~2 of \cite{2018PhR...733....1D}). At higher orders in perturbations,
more such terms that are protected by the equivalence principle appear,
and the EFT likelihood will consistently capture those as well once extended
to higher order.

\subsection{Marginalizing over bias parameters}
\label{sec:marg}

In the approach summarized in the previous section, a number of
nuisance parameters have been introduced to capture the uncertainties
associated with some of the poorly understood physical processes
describing halo and galaxy formation. In practical applications, these
parameters would then be estimated alongside cosmological parameters
in a statistical analysis of observational data. Interestingly, owing
to the simple functional form of the likelihood in \eq{eq:PcondGFT},
it is easily possible to analytically marginalize over some of the
bias parameters $b_O$. Here, we consider only those that do not appear
in the variance $\s^2(k)$; given \eq{eq:noise_var}, this includes all bias
parameters except $b_1$. While it might be possible to extend the analytical
marginalization to parameters which appear in the variance, the marginalization
performed here is sufficient for our purposes.

Let us thus write
\equ{
  \d_{h, \mathrm{det}}(\vk) = \mu(\vk) +
  \sum_{O\in \O_\text{marg}}
  b_O O(\vk) \, ,\qquad
  \mu(\vk) =
  \sum_{O \in \O_\text{all} \setminus \O_\text{marg}}
  b_O O(\vk) \, ,
}
where $\O_\text{marg}$ denotes the subset of operators, whose bias
parameters we wish to marginalize over (we denote the cardinality of
this set as $n_\text{marg}$). We can then write the likelihood
\refeq{PcondGFT} as
\ba
  P\left( \vec{\d}_h \Big| \vec{\d}, \{b_O\} \right) =\:& \frac1{{\cal N}}
  \exp \left[ -\frac{1}{2} \fsum{\vk} \ln \s^2(k)\right]
  \label{eq:PcondGFTmarg}\\
  &\times \exp \bigg\{ - \fsum{\vk}\bigg[ \frac{\left| \d_h(\vk) - \mu(\vk) \right|^2}{2\s^2(k)}
    - 2  \sum_{O\in \O_\text{marg}} b_O \frac{\Re\: [\d_h(\vk) -\mu(\vk)] O^*(\vk)}{2\s^2(k)} \vs
    & \hspace*{2.5cm}
    +   \sum_{O,O'\in \O_\text{marg}} b_O b_{O'} \frac{O(\vk)
    O'^*(\vk)}{2\s^2(k)} \bigg] \bigg\} \, ,
  \nonumber
\ea
where ${\cal N}$ is a normalization constant which is independent of
all parameters. This expression can be more compactly written as
\eqm[eq:Ppremarg]{
  P\left(\vec{\d}_h\Big|\vec{\d}, \{b_O\}\right) = \frac{1}{{\cal N}}
  \exp\left[-\frac{1}{2} \fsum{\vk} \ln \s^2(k)\right] \\
  \times \exp\left\{ - \frac{1}{2} C +   \sum_{O\in \O_\text{marg}} b_O B_O
  -\frac{1}{2}   \sum_{O,O'\in \O_\text{marg}} b_O b_{O'} A_{OO'} \right\} \, ,
}
where
\eqa[eq:margdefs]{
  C &= \fsum{\vk} \frac{1}{\s^2(k)} \left| \d_h(\vk) - \mu(\vk) \right|^2 \nn
  B_O &= \fsum{\vk} \frac{\Re\: [\d_h(\vk) -\mu(\vk)] O^*(\vk)}{\s^2(k)} \nn
  A_{OO'} &= \fsum{\vk} \frac{O(\vk) O'^*(\vk)}{\s^2(k)} \, .
}
Note that $A_{OO'}$ is a Hermitian and positive-definite matrix. The
former is immediately obvious from its definition. The latter follows
from the fact that $A_{OO'}$ is the zero-lag covariance matrix of a
set of sharp-$k$-filtered real fields $O(\vx)$. \refeq{Ppremarg} then
allows us to perform the Gaussian integral over the $b_O$. Here, we
will assume uninformative priors, although Gaussian priors can
trivially be introduced by adding a prior covariance to $A_{OO'}$. The
result is%
\footnote{Using the well-known Gaussian integral identity
  \equ{
    \int d^n \vec{x} \exp \left[ -\frac{1}{2} \vec{x}^\top A \vec{x} +
      \vec{B}^\top \cdot \vec{x} \right] = \frac{(2\pi)^{n/2}}{|A|^{1/2}}
    \exp \left[ \frac{1}{2} \vec{B}^\top A^{-1} \vec{B} \right] \, . \nonumber
  }
}
\eqa[eq:Pmarg1]{
  P\left(\vec{\d}_h\Big|\vec{\d}, \{b_O\}_\text{unmarg} \right)
  &= \left( \prod_{O\in \O_\text{marg}} \int db_O\right) P \left(
  \vec{\d}_h \Big| \vec{\d}, \{b_O\} \right) \nn
  & = \frac{(2\pi)^{n_\text{marg}/2}}{{\cal N}} \Big| A_{O O'} \Big|^{-1/2}
  \exp \left[ -\frac{1}{2} \fsum{\vk} \ln \s^2(k) \right] \nn
  & \times \exp \left\{ -\frac{1}{2} C(\{b_O\}) + \frac{1}{2}
    \sum_{O,O'\in \O_\text{marg}} B_O(\{b_O\}) (A^{-1})_{O O'}
  B_{O'}(\{b_O\}) \right\} \, .
}

We have thus reduced the parameter space from $\{b_O\}$ to
$\{b_O\}_\text{unmarg}$. This marginalization applies to an arbitrary number
of bias coefficients to be marginalized over. 
The price to pay is that we now need to
compute the vector $B_O$ and the matrix $A_{OO'}$. Further, we need
the determinant of $A_{OO'}$ and its inverse. Note, however, that
$n_\text{marg}$ -- and therefore the size of $B_{O}$ and $A_{O O'}$ --
will rarely become a large number in practical applications. More
specifically, in the tests presented below, we marginalize over
$\clapl,b_2,b_{K^2}$ and so $n_\text{marg}=3$. Further, $A_{OO'}$ is
independent of the remaining, unmarginalized bias
parameters.
On the other hand, $A_{OO'}$ does depend on the parameters
entering the variance $\s^2(k)$, \eq{eq:noise_var}.

At this point, let us briefly comment on the relation of this
analytic marginalization to other approaches presented in the recent
literature. In particular, Refs.~\cite{abidi/baldauf:2018,schmittfull/etal:2018} perform a Gram-Schmidt orthogonalization
on the fields entering the bias expansion \refeq{dhdet}, arguing that
lower-order operators thus become independent of higher-order fields
and make their corresponding bias parameters more robust to higher-order
corrections. This approach is directly related to the analytic marginalization
pointed out here. To see this, consider the case where an orthogonalization
has been performed on the operators. In particular, this implies that
$\<\mu(\vk)O(\vk)\>=0$ for all $O\in\O_\text{marg}$. This in turn renders
$B_O$ independent of all unmarginalized bias parameters, so that it becomes
a constant vector. Then, the marginalized likelihood reduces to the same form
as the unmarginalized likelihood keeping only the terms involving
unmarginalized bias parameters. In this sense, orthogonalization is
equivalent to marginalizing over bias parameters. The computational cost
of both approaches is expected to be essentially the same. However, the
marginalization described here does offer the possibility of including prior
information on the bias parameters that are marginalized over.

\subsection{Estimating systematic errors}
\label{sec:systematics}

One of the crucial advantages of the rigorous perturbative approach pursued
here is that it allows for an estimate of the systematic error due to
imperfections in the likelihood. We can distinguish three principal sources of such systematic errors (see \cite{cabass/schmidt:2019} for more details on Types 2 and 3 in particular):
\begin{description}
\item[Type 1:] Errors in the forward model for the matter density field (and correspondingly the operators constructed from it);
\item[Type 2:] Higher-order bias terms neglected in the expansion in $\d_{h,\rm det}$;
\item[Type 3:] Higher-order contributions to the variance $\s^2(k)$ as well as in  the form of the likelihood itself. 
\end{description}

The most rigorous way to evaluate the size of these contributions is to
include the set of leading higher-order terms that have been neglected
in the forward model, bias expansion, and likelihood, and evaluate the shift in
resulting parameter values. In case of the forward model (Type~1), this can be
tested by using the density field from N-body simulations instead of 2LPT
to construct the bias operators. This will be presented in \refsec{results}. 
In case of the bias expansion (Type~2), this test is not too difficult either, since the coefficients can be marginalized
over analytically, as shown in the previous section. We defer an implementation
of the higher-order contributions to future work however.

Let us here approximately estimate the size and scaling with $k_{\rm max}$ of
the systematic error of Type~2. Note that strictly speaking we have two cutoffs: the cutoff $\Lambda$ of the sharp-$k$ filter, and $k_{\rm max} < \Lambda$. In practice, one will choose $k_{\rm max}$ as a fixed fraction of $\Lambda$ (see \refsec{results}); hence, it is sufficient to consider the dependence on $\Lambda$ here.
For simplicity, we evaluate the systematic shift in the
bias parameters $b_O$. As described in \cite{paperI}, one can similarly evaluate
the shift in $\s_8$ by introducing scaled bias parameters $\beta_O$.
We will also count higher-derivative terms as higher order in perturbations, which assumes that the scale controlling higher-derivative terms does not differ greatly from the nonlinear scale. Thus, ``higher-order contributions'' include higher-derivative contributions in what follows.

Incorporating both sources of error described above,
the \emph{correct} likelihood can be written as
\ba
-2\ln P(\d_h | \d) &= \fsum{\vk}\left[\ln \s^2(k) + \frac1{\s^2(k)} \left| \d_h(\vk) + \d_h^\text{h.o.}(\vk) - \sum_O b_O \left(O + O_\text{err}\right)(\vk)\right|^2\right] \vs
&= - 2\ln P_\text{fid}(\d_h | \d) -2\sum_O B^\text{err}_O b_O + \sum_{O,O'} A^\text{err}_{OO'} b_O b_{O'} \,,
\label{eq:likeerr}
\ea
where
\ba
B^\text{err}_O &= \fsum{\vk} \frac1{\s^2(k)} \Re\left[  O_\text{err}(\vk) \d_h^*(\vk')' +  O^*(\vk) \d_h^\text{h.o.}(\vk')'\right] \vs
A^\text{err}_{OO'} &= \fsum{\vk} \frac1{\s^2(k)}  O_\text{err}(\vk) O'{}^*_\text{err}(\vk') '\,.
\label{eq:errdefs}
\ea
Here, $O_\text{err}(\vk)$ denotes the error field in the
operator $O$ due to deficiencies in the forward model, while
$\d_h^\text{h.o.}(\vk)$ denotes the higher-order bias contributions to
the actual halo density field. Finally, $P_\text{fid}(\d_h | \d)$ stands for the
fiducial likelihood, which differs from the correct one due to the systematic error terms. In the second line of \refeq{likeerr}, we have dropped an irrelevant constant term which does not depend on the parameters being varied. Let us write the fiducial likelihood in analogy to \refeq{Ppremarg} as
\ba
-2\ln P_\text{fid}(\d_h|\d) &= \fsum{\vk} \ln \s^2(k) + \sum_{O,O'} b_O b_{O'}\ohat{A}_{OO'} -2 \sum_O \ohat{B}_O b_O\,,\quad\mbox{where} \vs
\ohat{A}_{OO'} &= \fsum{\vk} \frac1{\s^2(k)}  O(\vk) O'{}^*(\vk') ' \vs
\ohat{B}_{O} &= \fsum{\vk} \frac1{\s^2(k)} \Re O(\vk) \d_h^*(\vk') ' \,.
\label{eq:fiddefs}
\ea
Under the assumption that the parameter shift $\Delta b_O$ due to the systematic errors is small, one can immediately solve for this shift based on the maximum-likelihood points of the correct and fiducial likelihoods.
  One can then estimate the expected amplitude of the shift by taking the expectation values of $\ohat{\v{A}}, \ohat{\v{B}}, \v{A}^\text{err}, \v{B}^\text{err}$. 
This is closely analogous to the ``Fisher bias.'' Using bold-face to denote vectors in the $n_O$-dimensional vector space of operators considered, we obtain in matrix notation
\ba
\Delta\v{b} \equiv \v{b} - \v{b}^\text{fid}
= \left( \ohat{\v{A}}^{-1} \v{A}^\text{err}\right) \left(\v{b}^\text{fid}\right) - \ohat{\v{A}}^{-1} \left(\v{B}^\text{err}\right)\,.
\ea
This expression involves the expectation value of the correlators in \refeq{errdefs} and \refeq{fiddefs} which are straightforward to evaluate in perturbation theory. We begin by estimating at which order in perturbation theory the various correlators contribute.

First, the expectation values of $\ohat{A}_{OO'}$ and $\ohat{B}_O$ are of order $\Plin(k) + \Pol(k)$ (in case of $\ohat{A}_{\d\d}$ and $\ohat{B}_\d$), or of order $\Pol(k)$ (all other elements). On the other hand, both $\v{B}^\text{err}$ and $\v{A}^\text{err}$ are of order $\Ptl(k)$. To see this, notice that both $O^\text{err}$ and $\d_h^\text{h.o.}$ are at least of cubic order in the linear density field. This means that all correlators which involve these error fields are two-loop contributions, apart from the cross-correlation with $\d^{(1)}$, which is at 1-loop order. The latter however only appears in $B^\text{err}_\d$, via $\< \d^\text{err,(3)} \d_h^{(1)}\>$ and $\< \d_h^{(3)} \d^{(1)}\>$. As we argue in App.~C of \citep{paperI}, these particular 1-loop contributions are of very similar shape as that coming from the higher-derivative bias, and are thus largely absorbed by $\clapl$.

Thus, without performing any detailed calculation, we can very roughly estimate that
\be
\Delta b \Big|_\text{loops} \sim \frac{\fsum{\vk} \s^{-2}(k) \Ptl(k)}{\fsum{\vk} \s^{-2}(k) \Plin(k)} \,.
\label{eq:syserror}
\ee
As an approximate estimate of the size of two-loop correlators, we will use the auto-correlation of $[\d^3]$:
\be
\Ptl(k) \sim \< [\d^3](\vk) [\d^3](\vk')\>' = 6\int d^3r\,[\xi_{\rm L}(r)]^3 e^{i\vk\cdot\v{r}}\,,
\label{eq:Ptl}
\ee
where $\xi_{\rm L}(r)$ is the linear matter correlation function.
We emphasize that this is a very rough estimate: in reality, $\Delta\v{b}$ involves many different contributions with various order-unity coefficients, which could add up or partially cancel. The main prediction of \refeq{syserror} is the scaling with $k_{\rm max}$.

Finally, let us consider systematics of Type~3, i.e. higher-order terms in the likelihood itself. Similar to the bias expansion, these come in two forms: an expansion in powers of $k$, equivalent to spatial derivatives; and an expansion in powers of perturbations, in this case the error field $\eps(\vk)$ whose variance is $\s_\eps^2$. Beginning with the former, a naive counting following loop contributions to the power spectrum indicates that a term
$C k^2$, where $C$ is a constant and which corresponds to the term $\propto \s_{\eps} \s_{\eps,2}$ in \refeq{noise_var}, is of 2-loop order (see, e.g., Sec.~4.1. of \citep{2018PhR...733....1D}). Hence, terms of order $C' k^4$ should have a negligible impact. This is corroborated by our numerical results (\refsec{results}).
The second type of higher-order stochasticity corresponds to non-Gaussian corrections such as the stochastic three-point function $\< \eps \eps \eps\>$ as well as coupling between stochasticity and the long-wavelength perturbations. These are briefly discussed in Sec.~5 of \citep{paperI} (see also \cite{cabass/schmidt:2019}).
A derivation of the precise form of these contributions to the likelihood requires more theoretical investigation, and is left for future work. However, we can  guess the approximate magnitude of these contributions by relating them to the terms we have kept here, which are nonlinear in the long-wavelength modes that determine $\d_\L$. The higher-order stochastic contributions are expected to be suppressed at least by 
\be
\Delta b \Big|_\text{stoch.} \sim \frac{|\eps(\vk)|}{|\d(\vk)|}\Big|_{k_{\rm max}} \sim \sqrt{\frac{P_\eps}{\Plin(k_{\rm max})}} \stackrel{\mbox{\small Poisson}}{=}
\Big[\bar{n} \Plin(k_{\rm max})\Big]^{-1/2}\,,
\label{eq:syserror2}
\ee
where in the last equality we have assumed the Poisson expectation, $P_\eps = 1/\bar{n}$ where $\bar{n}$ is the mean number density of halos, which is a reasonable first-order estimate for this purpose. While the proper
result will involve a summation over $\vk$ modes similar to \refeq{syserror},
we conservatively evaluate the ratio at $k_{\rm max}$ here, as it is unclear
what precise weighting should be employed for this type of higher-order contribution. Notice that \refeq{syserror2} also approaches 0 as $k\to 0$, but depends
sensitively on the abundance of halos. In particular, it becomes large for
small number densities.

\subsection{The \texorpdfstring{$\bm{\s_8}$}{sigma8} profile likelihood}
\label{sec:s8profile}

We now introduce the framework used in our numerical tests to obtain maximum
likelihood values and confidence intervals for cosmological parameters.
Below, we center our
discussion around the normalization of the primordial power spectrum,
described by \se.
The reasoning for choosing \se as parameter is that it can only be inferred by using information in the nonlinear density field, as mentioned in \refsec{likeF}. An unbiased inference thus means that the specific part of the information content in the nonlinear density field that is robust has been properly isolated.
In particular, nonlinear information is explicitly necessary in order
to break the degeneracy with the bias parameter $b_1$, rendering it
the most direct test of our nonlinear inference approach. Future work
will consider other cosmological parameters as well.

Besides from being a function of \se, the marginalized likelihood
\eq{eq:Pmarg1} also depends on bias and other nuisance parameters
(including the entire set of Fourier modes of the three-dimensional matter density field). Since the
probabilistic inference of the initial matter density field from tracers like
a halo catalog is numerically very expensive \cite[see,
  e.g.,][]{2013MNRAS.432..894J}, we instead constrain it to the actual
initial conditions used in the simulations, evolved to low redshifts
using either second-order Lagrangian perturbation theory (2LPT), or
the N-body code directly. This forward evolution is then performed for
a set of discrete $\se$ values around the fiducial $\se=0.85$
(see \refsec{impl}). We can then maximize the
likelihood to simultaneously obtain best-fit values for cosmological
and the remaining nuisance parameters.  As mentioned in
\refsec{intro}, this is in fact the most stringent test possible for any systematic bias,
since the absence of any flexibility in the phases means that there is less room for
errors in the likelihood to be absorbed by changing the initial
conditions.

On the other hand, by fixing the phases, the only way to obtain
rigours error estimates would be to analyze a large ensemble of
large-volume simulations. Since these are costly, we here resort to a
different method, allowing us to obtain error estimates from the
likelihood itself: the profile likelihood method \cite{wilks1938}
provides estimates of confidence intervals for individual parameters
of multivariate distributions within a frequentist approach. For a
probability distribution $P(a, \{b_i\})$, the profile likelihood for
parameter $a$ is defined as
\equ[eq:prof_def]{
  \pprof(a) = \underset{\{b_i\}}{\mathrm{arg \, max}} [ P(a, \{b_i\}) ] \, ,
}
where the additional set of parameters $\{b_i\}$ has been profiled
out.
Constructing a full profile likelihood for \se is still numerically
expensive, since it formally requires a recomputation of the final
matter density field each time the function argument is updated. To
speed up the analysis, we instead interpolate the profile likelihood evaluated on a
predefined grid in \se centered about the fiducial value of the
simulation. The details of this procedure will be
described in \sect{sec:impl}.

\section{Simulations}
\label{sec:sim}

All numerical tests presented below are based on the N-body simulations
presented in \cite{2017MNRAS.468.3277B}. They are generated using
GADGET-2 \cite{2005MNRAS.364.1105S} for a flat $\Lambda$CDM cosmology
with parameters $\Omega_\mathrm{m} = 0.3$, $n_\mathrm{s} = 0.967$, $h = 0.7$,
and $\se = 0.85$, a box size of $L = 2000 \, h^{-1}
\mathrm{Mpc}$, and $1536^3$ dark matter particles of mass
$M_\mathrm{part} = 1.8 \times 10^{11} \, h^{-1} M_{\odot}$. Initial
conditions for the N-body runs were obtained at redshift $z_\mathrm{ini}
= 99$ using second-order Lagrangian perturbation theory (2LPT)
\cite{1993MNRAS.264..375B} with the 2LPTic algorithm
\cite{2006MNRAS.373..369C, 2012PhRvD..85h3002S}.
Ref.~\cite{2017MNRAS.468.3277B} also presented runs for two further
values of \se bracketing the fiducial value in order to perform the
derivative of the halo mass function with respect to \se (for studies
related to the scale-dependent bias induced by primordial non-Gaussianity).
We use these, in addition to simulations for 4 additional \se values which
  were performed specifically for this paper, as well.

Dark matter halos
were subsequently identified at redshift $z = 0$ as spherical
overdensities \cite{1974ApJ...187..425P, 1992ApJ...399..405W,
  1994MNRAS.271..676L} applying the Amiga Halo Finder algorithm
\cite{2004MNRAS.351..399G, 2009ApJS..182..608K}, where we chose an
overdensity threshold of 200 times the background matter density.
The halo samples considered here consist of two mass ranges each at
two redshifts, and are summarized in \reftab{halos}.
We only consider halos above $10^{13} \Msunh$, corresponding to a minimum
of 55 member particles.
Note that differences in the number densities of the halo samples imply
differences in the expected parameters as well as errors on the inferred \se, an important aspect in our validation of the inference
framework.

\begin{table}[b]
\centering
\begin{tabular}{c c c c c c}
\hline
\hline
Redshift & \specialcell{Mass range\\$\log_{10} M [\Msunh]$} & $N_\text{halo}$ (run 1) & $N_\text{halo}$ (run 2) & $\bar n\ [(\Mpch)^{-3}]$ \\
\hline
0 & $[13, 13.5]$ & 2807757 & 2803575 & $3.5 \times 10^{-4}$ \\
0 & $[13.5, 14]$ & 919856 & 918460 & $1.1 \times 10^{-4}$ \\
\hline
1 & $[13, 13.5]$ & 1507600 & 1506411 & $1.9 \times 10^{-4}$ \\
1 & $[13.5, 14]$ & 301409 & 302182 & $3.8 \times 10^{-5}$ \\
\hline
\hline
\end{tabular}
\caption{The halo samples used in our numerical tests. Throughout, masses $M \equiv M_{200m}$ are spherical-overdensity masses with respect to 200 times the background matter density.}
\label{tab:halos}
\end{table}

\section{Implementation}
\label{sec:impl}

We now provide additional details of the setup and numerical
implementation used in our tests. We take as given the halo catalogs
described in the previous section, as well as a set of matter particles generated
either via 2LPT or full N-body for a set of values.
Since the matter density field is a function of \se, our main
parameter of interest, mapping the profile likelihood as a continuous
function of \se would require to recompute the set of operators for
each function evaluation. To expedite the analysis, we instead
generate representations of the evolved matter density field
for fixed initial phases and a discrete set of values for \se (as well as
redshifts $z=0$ and $z=1$ of the halo samples) given by
\ba
\se &\in \{ 0.65,\  0.75,\  0.80,\  0.83,\  0.85,\  0.87,\  0.90,\  0.95,\  1.00,\  1.10,\  1.20 \} \quad\mbox{(2LPT)}\,, \vs
\se &\in \{ 0.68,\  0.78,\  0.83,\  0.85,\  0.87,\  0.92,\  1.02 \} \quad\mbox{(N-body)}\,.
\ea

For a given halo sample at a given
redshift, and a fixed value $\se^i$, the steps for evaluating the \se profile likelihood are as
follows:
\begin{enumerate}
\item The halos and matter particles are assigned to a $1024^3$ grid using a cloud-in-cell density assignment. The high resolution is chosen to avoid leakage of the assignment kernel to the low wavenumbers of interest.
\item A sharp-$k$ filter is applied to the matter and halo fields in Fourier space, such that modes with $|\vk| > \Lambda$ are set to zero. The Fourier-space grids are subsequently restricted to $384^3$, chosen such that the Nyquist frequency of each grid remains above $3\Lambda$ for all values of $\Lambda$ considered here.
\item Quadratic and higher-derivative operators are constructed from the sharp-$k$ filtered matter density field and held in memory. The quadratic operators are renormalized following \app{app:operators}.
\item The maximum of the likelihood over the parameter space spanned by the remaining bias and stochastic variance parameters is then found via
  function minimizer MINUIT \cite{James:1975dr} (in practice we minimize $-2\ln\mathcal{L}$, i.e. the pseudo-$\chi^2$). The operator fields do not need to be recomputed for each evaluation, as only their coefficients are varied.
\end{enumerate}
More precisely, we employ the analytic marginalization described in \sect{sec:marg} for the parameters $b_2$, $\clapl$ and $b_{K^2}$, leaving only $b_1$
and the three stochastic amplitudes to be varied in the minimization.
We have not found any significant impact of the term $\s_{\eps\eps_m,2}$
in \eq{eq:noise_var}, but a significant degeneracy with $\s_{\eps,2}$.
For this reason, we fix the former to zero in our default analysis.
This leaves a three-dimensional parameter space to be searched in the
minimization, which typically converges quickly. We have found the minimization
robust to varying initialization points and number of successive MINUIT
cycles. Our default choice for the maximum wavenumber in the likelihood
$k_\mathrm{max}$ is $\Lambda/2$. 

This procedure results in a set of values $\{ \se^i,\, -2\ln P^\text{prof}(\se^i) \}_i$
which we find is fit well by a parabola in all cases (we disregard values of \se where the minimization failed to converge).
The best-fit value $\hse$ is given by the minimum point of the
best-fit parabola, while the estimated $1\sigma$ error on $\hse$ is given by the inverse square-root of the curvature of the parabolic fit. 
We emphasize that this error does not include any residual cosmic variance,
and is essentially purely governed by the halo stochasticity which appears in
the variance of the likelihood. 

  \twofig{{{plots_paper/prof_run1_L0.10_z0_lM13.0-13.5}}}
           {{{plots_paper/prof_run1_L0.10_z0_lM13.5-14.0}}}
           {fig:prof}
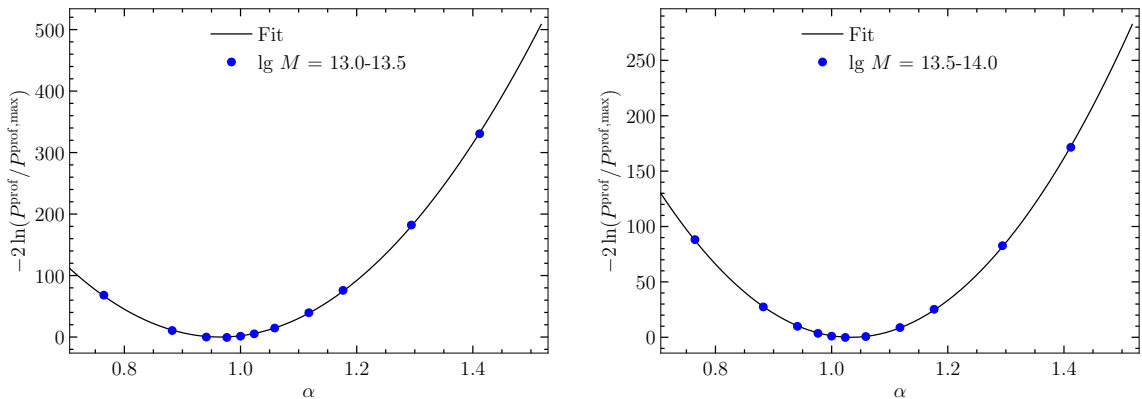
           {Examples of the profile likelihood $-2\ln P^\text{prof}(\se)/P^\text{prof}(\hse)$, plotted as a function of $\alpha \equiv \se/\se^\text{fid}$. In all cases, results are for run 1 and $\Lambda = 0.1\iMpch$ at $z=0$. Also shown are the parabolic fits whose maximum point results in the value $\hse$ or equivalently $\hat{\alpha}$ listed in \reftab{results}. The value for $-2\ln P^\text{prof}(\hse)$, which is subtracted for better readibility, is taken from the parabolic fit.}

\section{Results}
\label{sec:results}

In the following, we present results for the best-fit value $\hse$. For
convenience, we phrase these in terms of the ratio to the fiducial value,
introducing
\be
\hat{\alpha} \equiv \frac{\hse}{\se^\text{fid}}.
\label{eq:alphadef}
\ee
First, \reffig{prof} shows examples of the profile likelihood determined as
described in the previous section, with the parabolic fit that is used to determine $\hse$. All panels are for $\Lambda = 0.1\iMpch$. Clearly, the log-profile likelihood is well approximated by a parabola, so that we expect maximum point and curvature to yield unbiased estimates of the maximum and 68\%-level confidence intervals with respect to a full scan of the profile likelihood. The results for all halo samples and $\Lambda=0.1\iMpch$, our fiducial choice, are summarized in \reftab{results}. We find that an unbiased value of $\hse$ is recovered to within $\sim 2\sigma$ in most cases.
Notice that the run-to-run variance is larger than the estimated error bars in several cases. This could be due to residual cosmic variance, which is not contained in the estimated error bars as discussed in the previous section, although possible issues with the minimizer in isolated cases also cannot be excluded. In order to investigate this, more realizations would be needed.

The remaining columns in \reftab{results} show the value of $b_1$ as well as the stochastic amplitudes, all corresponding to the maximum-likelihood point for the fiducial value $\se^\text{fid}$. Recall that all other parameters are analytically marginalized over. The bias $b_1$ is essentially fixed by the cross-correlation of $\d_h$ with $\d$. Correspondingly, we find that the combination $b_1 \se$ is constant for all \se values to within several percent. The stochastic amplitude $\s_\eps^2$ is scaled to the Poisson expectation following \refapp{variance}. 
Values greater (less) than one thus correspond to super- (sub-)Poisson stochasticity. We do find evidence for a smaller stochasticity for the rarer halo samples, in agreement with previous findings \cite{hamaus/etal:2010}. The last column shows the ratio of the higher-order (in $k^2$) stochastic parameter to the leading-order one. This gives a rough indication for the spatial length scale squared associated with the scale-dependent stochasticity. We thus find this length scale to be of order $(1-5)\Mpch$. Notice however that this parameter is expected to also absorb various higher-order contributions not explicitly included in the likelihood, as discussed in \citep{paperI}, so that one cannot robustly infer a physical length scale from this value.

\begin{figure*}[t]
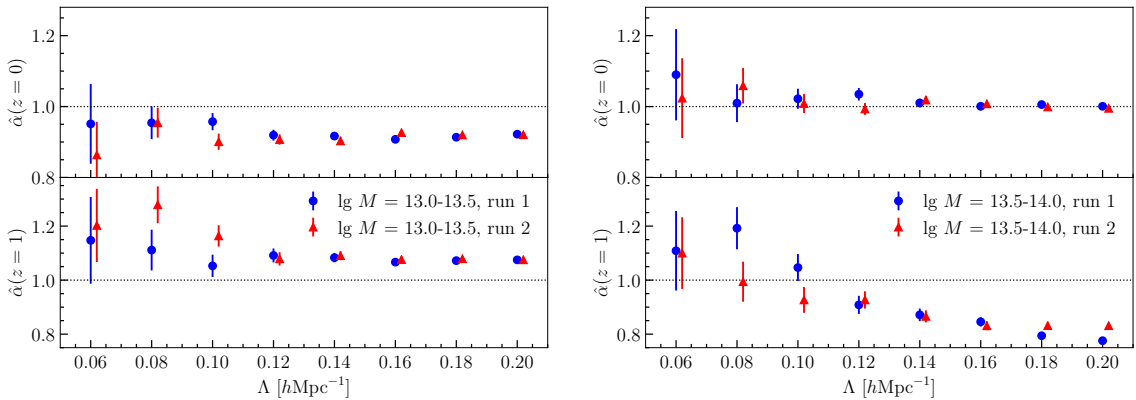
%
  \centerline{\resizebox{\hsize}{!}{
      \includegraphics*{{{plots_paper/sigma8_mlemargdef_s3fix_f_lM13.0-13.5}}}
      \includegraphics*{{{plots_paper/sigma8_mlemargdef_s3fix_f_lM13.5-14.0}}} }}
  \caption{Best-fit values $\hat\alpha = \hse/\se^\text{fid}$ as a function of the cutoff $\Lambda$ (with $k_\text{max} = \Lambda/2$ in each case) for the low- and intermediate-mass samples in the left and right panels, respectively. In each panel, the upper plot shows results at $z=0$ while the lower plot shows $z=1$.
    \label{fig:s8vLambda}}
\end{figure*}%

Allowing $\L$ to vary, we obtain the results shown in \reffig{s8vLambda}. 
In each case, we show results for both simulation runs.
           While the differences in $\hat{\alpha}$ from unity are broadly consistent with being residual stochasticity and cosmic variance for $\Lambda \leq 0.1 \iMpch$, this no longer holds for higher cutoff values which at $z=1$ should still be under good perturbative control.
           On the other hand, the results appear remarkably stable toward higher values of $\L$ up to $\L=0.2\iMpch$ in many cases. Notice that the majority of the modes contributing to the profile likelihood with $\L=0.14\iMpch$, say, are not included in the likelihood with $\L=0.1\iMpch$, so that these are largely independent measurements.

\begin{table}[b]
\centering
\begin{tabular}{c c c c c c c}
\hline
\hline
Redshift & \specialcell{Mass range\\$\log_{10} M [\Msunh]$}
& $\hat{\alpha}$ (run 1) & $\hat{\alpha}$ (run 2) & $b_1$
& \specialcell{$\sigma_\eps^2$\\$[$Poisson$]$}
& \specialcell{$\sigma_{\eps,2}/\sigma_\eps$\\ $[(\Mpch)^2]$} \\
\hline
0 & [13.0-13.5] & $0.96 \pm 0.02$ & $0.90 \pm 0.02$ & $1.20$ & $1.11$ & $-25.5$ \\
0 & [13.5-14.0] & $1.02 \pm 0.03$ & $1.01 \pm 0.03$ & $1.61$ & $0.96$ & $-11.7$ \\
\hline
1 & [13.0-13.5] & $1.05 \pm 0.04$ & $1.16 \pm 0.04$ & $2.36$ & $0.93$ & $1.3$ \\
1 & [13.5-14.0] & $1.05 \pm 0.05$ & $0.93 \pm 0.05$ & $3.49$ & $0.89$ & $10.5$ \\
\hline
\end{tabular}
\caption{Summary of results for $\Lambda = 0.1\iMpch$ and $k_{\rm max} = \Lambda/2$, with the likelihood settings described in the text. For the best-fit scaled \se estimate $\hat{\alpha}$, results from run 1 and run 2 are shown individually with estimated 68\% CL error bars. $b_1$ and stochastic amplitudes are reported for the fiducial $\se=\se^\text{fid}$ and averaged over both runs. The stochastic variance $\sigma_\eps^2$ is scaled to the Poisson expectation for the given halo sample, as described in \app{app:variance}. The last column shows the ratio of the higher-derivative stochastic amplitude to the lowest-order one, indicating the scale associated with the expansion of $\s^2(k)$ in $k$.}
\label{tab:results}
\end{table}

Before turning to possible explanations for these trends, let us briefly comment on the choice of $k_\text{max}/\Lambda$. We do not find strong trends with this parameter. Increasing $k_\text{max}$ at fixed $\Lambda$ yields similar trends as increasing $\Lambda$ itself, which is shown in \reffig{s8vLambda}. For this reason, we fix $k_\text{max}=\Lambda/2$ throughout.

\begin{figure*}[t]
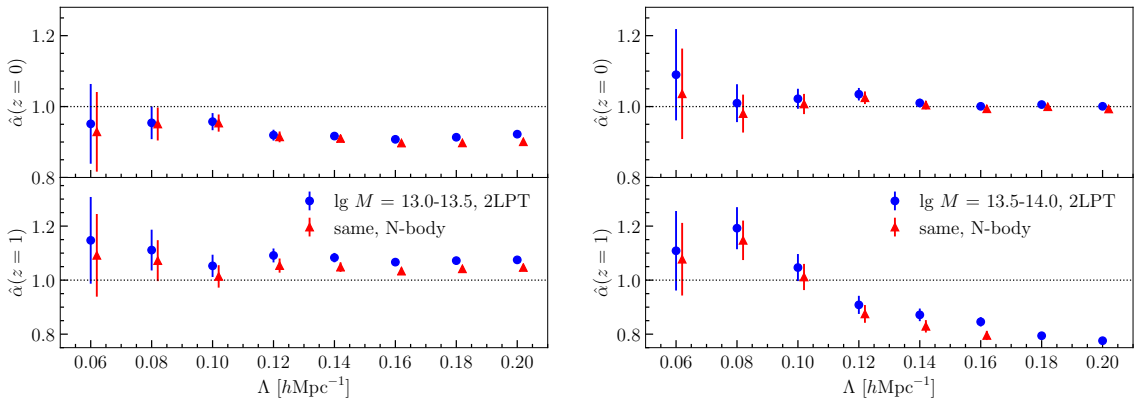
%
  \centerline{\resizebox{\hsize}{!}{
      \includegraphics*{{{plots_paper/sigma8_mlemargNbody_s3fix_f_lM13.0-13.5}}}
      \includegraphics*{{{plots_paper/sigma8_mlemargNbody_s3fix_f_lM13.5-14.0}}} }}
  \caption{Same as \reffig{s8vLambda}, but using the density field from the N-body simulation. Results for run 1 are shown. 
    \label{fig:s8vLambdaNbody}}
\end{figure*}

As a test of the systematics of Type~1, we use the N-body density field itself instead of 2LPT for the construction of the bias operators. The result is shown in \reffig{s8vLambdaNbody}. We only find minor shifts in $\hse$. Indeed, the cross-correlation coefficient between the 2LPT and N-body density fields in Fourier space is better than 0.97 for all scales and redshifts considered here, so a large shift would be surprising. Although there is some improvement, we conclude that the 2LPT density field is not primarily responsible for the bias in $\hse$ found.

We next turn to systematics of Type~3, specifically the possible impact of the chosen implementation of $\s^2(k)$ in terms of two free parameters (since we fix $\s_{\eps\eps_m,2}$ to zero). Since the halo number density is smaller at higher redshift, and the stochasticity correspondingly larger, this could possibly explain the increased bias in \hse at $z=1$ compared to $z=0$. 
Performing the profile likelihood analysis with varying $\s_{\eps\eps_m,2}$ on the one hand, and both the former parameter and $\s_{\eps,2}$ fixed to zero on the other, leads to sub-percent shifts in \hse (this can also be gleaned by the values shown in the right-most column of \reftab{results}, which, when multiplied by $k_\text{max}^2$ indicate the upper bound on the fractional contribution of the term $\s_{\eps,2}k^2$ to the total variance). Thus, we conclude that the parametrization of $\s^2(k)$ is unlikely to be responsible for the bias in \hse as well.

This leaves two possible sources of systematic error: our Type-2 systematic, i.e. higher-order terms in the bias expansion (both in perturbations and derivatives); and other systematics of Type-3, namely higher-order corrections to the form of the likelihood itself.
Both types of terms are expected to be largest for the rarest and most highly biased halo samples. 
It is worth nothing that higher-order bias contributions are not necessarily smaller at higher redshift for fixed halo mass, since higher-redshift samples are more biased. Further, higher-derivative terms, which are expected to be tied to the Lagrangian radius of halos, most likely do not decrease toward higher redshift (see for eample the results of \cite{lazeyras/schmidt}). Indeed, preliminary investigations show that incorporating higher-order terms in the derivative expansion, such as $(\lapl)^2\d$ and $\lapl(\d^2)$, in the set of operators does have an impact on the profile likelihood. We leave a detailed investigation of the impact of higher-order bias terms to upcoming work.

The remaining Type-3 systematics are expected to be controlled by the ratio in \refeq{syserror2}; this turns out to be of order unity or larger for the halo samples considered here, implying that higher-order stochastic corrections could be significant. At this point, lacking an explicit expression for these terms, it is difficult to quantitatively evaluate their impact however.

  Additional investigations have pointed to a likely cause for the bias in the inferred value of $\sigma_8$ which is related to a higher-order term that is enhanced on large scales. Specifically, the auto-correlation of the quadratic bias operators $A_{OO'}$ contains a contribution from the connected matter four-point function (trispectrum), which, for the likelihood and second-order bias expansion used here, corresponds to an error term $A_{OO'}^\text{err}$ (\refsec{systematics}). 
  For small $k$, the dominant term from the particular trispectrum configuration involved scales as $\< \d_\L^2\>^2 P_{\rm L}(k)$, where $P_{\rm L}(k)$ is the linear matter power spectrum. While this contribution is suppressed compared to the leading, disconnected term, the latter approaches a constant at small $k$, while the trispectrum contribution grows toward small $k$ (assuming $k\gtrsim 0.02 \iMpch$) due to the factor $P_{\rm L}(k)$. For this reason, it can bias the maximum-likelihood value for $\sigma_8$ even for very small values of $\Lambda$ which correspondingly push $k$ to small values (see the behavior of $\alpha$ for $\Lambda \lesssim 0.1 \iMpch$ in \reffig{s8vLambda}). We leave a more detailed investigation, and possible remedies, for future work.

\section{Conclusions}
\label{sec:concl}

We have presented the results of a first application of the effective-field-theory-based Fourier-space likelihood derived in \citep{paperI} to halo catalogs. More precisely, the test case is to obtain an unbiased estimate of the amplitude of the linear power spectrum \se (or equivalently normalization of scalar perturbations $\mathcal{A}_s$) purely based on the nonlinear information in the halo density field that is protected by the equivalence principle. For this, we vary four bias parameters as well as two stochastic amplitudes. The reasoning for choosing \se as parameter is that it can only be inferred by using information in the nonlinear density field. An unbiased inference---which we have not completely achieved yet---would thus mean that the robust, protected nonlinear information content has been isolated. We expect that other parameters, such as the BAO scale or the matter power spectrum shape, will then also be unbiased, as they rely to a lesser degree on purely nonlinear information.

We further presented a method to analytically marginalize over bias parameters, which we apply to three of the four bias parameters in our implementation. 
We expect that this analytical marginalization will prove extremely powerful when going to higher order in the bias expansion (both in orders of perturbations and derivatives): when using this technique, the cost of finding the maximum-likelihood point, or more generally, sampling from the likelihood, only increases quadratically with the number of bias terms (since the cost is dominated by the evaluation of the matrix $A_{OO'}$).\footnote{This scaling is based on the fact that the computational time of evaluating the likelihood is dominated by the determination of the quantities $A_{OO'}$ and $B_O$ in the notation of \refsec{marg}. The cost of the matrix inversion is negligible for a realistic number of operators.} On the other hand, the computational cost would grow much more rapidly if one where to explicitly vary all bias parameters. 

Our results indicate that \se can be recovered with a systematic error under $\sim 10\%$ for a range of halo samples at different redshifts when using a cutoff value $\L=0.1 \iMpch$.
Our assessment is that the most likely explanation for the residual systematic bias in \se can be traced to a trispectrum contribution to the quadratic operator correlators. Precisely how this contribution can be taken into account via higher-order terms in the bias expansion or likelihood is left for future work.
It would further be interesting to perform a joint inference from the different halo samples that we have considered separately here, which however requires a generalization of the likelihood to include the stochasticity cross-covariance between different tracers. We leave these developments to future work.

Our approach has some resemblance to what was recently presented in \citep{schmittfull/etal:2018}. Instead of the sharp-$k$ filter employed here (necessary for an unbiased inference following \citep{paperI}), the authors of \cite{schmittfull/etal:2018} used a Gaussian filter. More importantly, they allowd for the bias coefficients to be free functions of $k$, $b_O \to b_O(k)$, which removes the information on \se, and instead focused on the degree of cross-correlation of the field we call $\d_{h,\rm det}$ here with the actual halo density field. It would be interesting to study the corresponding correlation coefficient for our, sharp-$k$-filtered field $\d_{h,\rm det}$ given the best-fit bias parameters. We defer this to future work as well. 

A natural next question is: what is the expected statistical error for the inferred \se in the realistic case when the phases of the linear density field are unknown? In order to determine this, one unfortunately has to marginalize over those phases, which requires an implementation of the EFT likelihood into a sampling framework along the lines of, e.g. \cite{2013MNRAS.432..894J}. We thus have to defer this question to future work as well. It is clear however that this uncertainty will be very sensitive to the value of the cutoff $\L$.  
Regardless of the expected statistical precision, it is worth emphasizing that, by fixing the phases to their ground-truth values throughout, the test of unbiased cosmology inference presented here is the most stringent test possible.

Finally, in parallel to the empirical studies on halo samples, a more rigorous theoretical study of the EFT likelihood expansion should be performed \cite{cabass/schmidt:2019}. This is essential in order to obtain proper estimates of the relative size of the different expansion parameters, and to consistently carry out the expansion to higher order. After all, one of the main advantages of the EFT likelihood approach is that it is very simple to systematically go to higher orders.

\acknowledgments

We thank Alexandre Barreira, Giovanni Cabass, Stefan Hilbert, Marcel Schmittfull, Marko Simonovi\'c, and Matias Zaldarriaga for helpful discussions. We further thank Titouan Lazeyras for supplying us with the N-body and 2LPT codes used to generate initial conditions for the simulations our results are based on. 
FE, MN, and FS acknowledge support from the Starting Grant (ERC-2015-STG 678652) ``GrInflaGal'' of the European Research Council.
GL acknowledges financial support from the ILP LABEX (under reference ANR-10-LABX-63) which is financed by French state funds managed by the ANR within the Investissements d'Avenir programme under reference ANR-11-IDEX-0004-02.  This work was supported by the ANR BIG4 project, grant ANR-16-CE23-0002 of the French Agence Nationale de la Recherche.
This work is done within the Aquila Consortium.\footnote{\url{https://aquila-consortium.org}}

\appendix

\section{Operator correlators and renormalization}
\label{app:operators}

As argued in \citep{paperI} (see also \cite{schmittfull/etal:2018,abidi/baldauf:2018}), nonlinear operators constructed out of the matter
density field must be renormalized in order
to suppress dependencies of their cross-correlations on small-scale
modes \citep{2014JCAP...08..056A}. Specifically, using square brackets
to denote renormalized operators $[O]$, we found using a tree-level (leading-order) calculation \citep{paperI}
\eqa[eq:renormfid]{
  [\d](\vk) &= \d(\vk) \nn
  \Bigl[ \d^2 \Bigr](\vk) &= (\d^2)(\vk) - \Sigma_{1-3}^2(k) \d(\vk)
  \quad \mbox{and} \quad [\d^2](\vk=0) = 0 \nn
  \Bigl[K^2 \Bigr](\vk) &= (K^2)(\vk) - \frac23 \Sigma_{1-3}^2(k) \d(\vk)
  \quad \mbox{and} \quad [K^2](\vk=0) = 0 \, ,
}
where
\equ[eq:sigma13def]{
  \Sigma_{1-3}^2(k) = 4\int_{\vp} W_\Lambda(\vp) W_\Lambda(\vk-\vp) F_2(-\vk,\vp) \Plin(p) \, .
}
Here, the $W_\Lambda(k)$ are sharp filter functions defined in Fourier
space. Finally,
\equ[eq:f2def]{
  F_2(\vk_1,\vk_2) = \frac{5}{7} + \frac{2}{7} \frac{(\vk_1 \cdot \vk_2)^2}{k_1^2 k_2^2}
  + \frac{\vk_1 \cdot \vk_2}{2k_1 k_2} \left( \frac{k_1}{k_2} + \frac{k_2}{k_1} \right) \, .
}
Since this calculation is only valid at leading order, we use a numerical
renormalization procedure in our likelihood implementation instead.
Specifically, we measure
\be
P_{\d O^{[2]}}(k) = \< \delta(\vk) O^{[2]}(\vk')\>'
\ee
on a linear grid in $k$ (we choose 100 bins between the fundamental and Nyquist frequencies). The same is done for the density field itself, yielding $P_{\d\d}(k)$. Then, for each mode $\vk$, we renormalize through
\be
[O^{[2]}](\vk) = O^{[2]}(\vk) - \frac{P_{\d O^{[2]}}(|\vk|)}{P_{\d\d}(|\vk|)} \d(\vk),
\ee
where a cubic-spline interpolation is used to obtain the power spectra
at each value of $\vk$. \reffig{opcorr1} shows the cross-correlation of $\d$ and the two quadratic operators $\d^2$ and $K^2$ before and after renormalization. For the $k$ values that matter most in the likelihood, $k \gtrsim 0.02 \iMpch$, the cross-correlation is removed to high accuracy by the renormalization procedure.
Also shown is the tree-level perturbation-theory prediction for the correlator before renormalization, which matches the measurement reasonably well, although not perfectly even at low $k$, since modes near the cutoff $\Lambda$ contribute to this cross-correlation.

\twofig{{{plots_paper/opcorr_dd2}}}
       {{{plots_paper/opcorr_dK2}}}
       {fig:opcorr1}
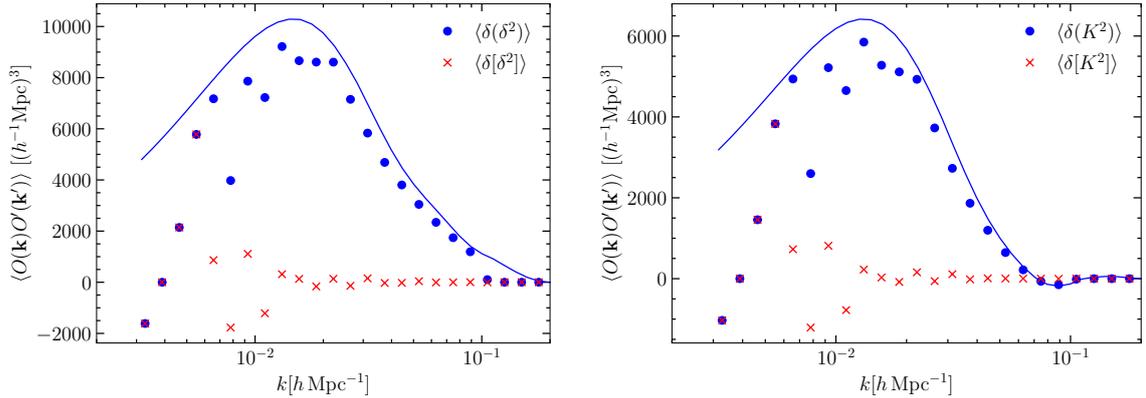
       {\textit{Left panel:} Measured operator correlator $\< \delta_\L (\delta_\L)^2 \>$ (i.e. before renormalization) and $\< \delta_\L [\delta_\L]^2 \>$ (after renormalization) using a 2LPT density field at $z=0$ with $\Lambda=0.1\iMpch$. The line shows the tree-level standard perturbation theory prediction.
       \textit{Right panel:} Same as left panel, but for $\< \delta_\L (K_\L^2)\>$ and $\<\delta_L [K_\L]^2\>$.}

\reffig{opcorr2} shows the cross-correlation of the quadratic operators among each other. As argued in \citep{paperI}, the renormalization also removes the dominant higher-order (trispectrum) contribution to these correlators. Indeed, the leading perturbation-theory prediction matches the cross-correlation of the quadratic operators well. 
We have verified that the good agreement also holds for other values of redshift and $\L$, and that the deviations from the perturbation-theory prediction show the expected scaling, with agreement improving toward higher redshift and for smaller values of the cutoff $\L$. We conclude that the operator cross-correlations, which form the practical basis of the EFT likelihood as discussed in detail in \citep{paperI}, are well understood.

\onefigs{{{plots_paper/opcorr_O2O2}}}
       {fig:opcorr2}
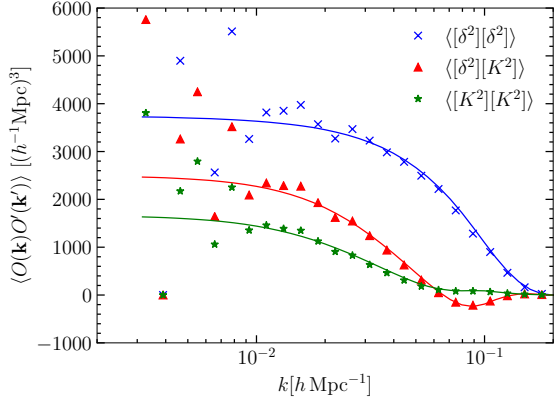
       {Measured renormalized operator correlators $\< [O^{[2]}] [O'^{[2]}]\>$ for $O,O' \in \{ (\delta_\L)^2, (K_\L)^2\}$. All operators are constructed from a 2LPT density field at $z=0$ with $\Lambda=0.1\iMpch$. The lines again show the tree-level standard perturbation theory predictions.}

\section{Interpreting the variance \texorpdfstring{$\bm{\s_\eps^2}$}{sigma_eps}}
\label{app:variance}

In this appendix we derive the expectation for the variance parameter $\sigma_\eps^2$ for Poisson noise. Neglecting long-wavelength perturbations, let $n_i \equiv n(\vx_i)$ denote the number density of halos in the grid cell centered around $\vx_i$. Assuming this is Poisson distributed, we obtain
\be
\lambda \equiv \< n_i\> = \frac{N_h}{N_g^3} \quad\mbox{and}\quad
\< n_i n_j \> = \lambda \d_{ij}\,,
\ee
where $N_h$ is the total number of halos in the box. The noise in the fractional halo density perturbation $\d_h$ is then given by $\eps_i = n_i / \lambda$, where we neglect the subtraction of the mean here since it is irrelevant for modes of finite $\vk$. The noise field obeys, under the Poisson assumption,
\be
\< \eps_i \eps_j \> = \frac1{\lambda} \d_{ij}\,.
\ee
Finally, its power spectrum is given by
\ba
\< |\eps(\vk)|^2 \> &= \sum_{i,j} \< \eps_i \eps_j \> e^{i\vk\cdot(\vx_i-\vx_j)} = \frac{N_g^3}{\lambda} = \frac{N_g^6}{\bar{n}_h \Lbox^3}\,,
\label{eq:Pkeps}
\ea
where $\bar{n}_h = N_h/\Lbox^3$ is the number density of halos. Notice that the value depends on the grid resolution adopted, which in our implementation is $N_g = 384$. The values of $\s_\eps^2$ given in \reftab{results} are divided by this Poisson expectation. Values greater (less) than one thus correspond to super- (sub-)Poisson stochasticity.

\bibliographystyle{JHEP}
\bibliography{bibliography}

\end{document}